# Flexible single multimode fiber imaging using white LED

Minyu Fan[1], Kun Liu[1], Jie Zhu[1], Yang Hu[1], Songsong Zhu[2], Nan Jiang[2], Sha Wang[1*]

[1] College of Electronics and Information Engineering, Sichuan University, Chengdu 610064, People's Republic of China
[2] State Key Laboratory of Oral Diseases & National Center for Stomatology & National Clinical Research Center for Oral Diseases & Department of Orthognathic and TMJ Surgery, Sichuan University, Chengdu 610064, People's Republic of China
E-mail: shawang@scu.edu.cn

**Abstract**

Multimode fiber (MMF) has been proven to have good potential in imaging and optical communication because of its advantages of small diameter and large mode numbers. However, due to the mode coupling and modal dispersion, it is very sensitive to environmental changes. Minor changes in the fiber shape can lead to difficulties in information reconstruction. Here, white light emitting diode (LED) and cascaded Unet are used to achieve MMF imaging to eliminate the effect of fiber perturbations. The output patterns in three different color channels of the charge-coupled device (CCD) camera produced by transferring images through the MMF are concatenated and inputted into the cascaded Unet using channel stitching technology to improve the reconstruction effects. The average pearson correlation coefficient (PCC) of the reconstructed images from the Fashion-MINIST dataset is 0.83. In order to check the flexibility of such a system, perturbation tests on the image reconstruction capability by changing the fiber shapes are conducted. The experimental results show that the MMF imaging system has good robustness properties, i. e. the average PCC remains 0.83 even after completely changing the shape of the MMF. This research potentially provides a flexible approach for the practical application of MMF imaging.

## 1. Introduction

Multimode fibers (MMFs) have attracted researchers' attention in the fields of optical communications [1, 2] and imaging [3-6] since they are slender and have a great number of modes that can transmit a large amount of information in parallel [1]. Due to the presence of mode coupling and modal dispersion within the MMF, speckles are generated when the light goes through the fiber [6]. Normally coherent light is proposed to behave as the light source [7-20]. Focusing and wide-field imaging after the MMF have been successfully achieved by using the methods such as phase conjugation [4, 7-12], transmission matrix [3, 13-17], and iteration algorithm [5, 18-20]. However, these methods require maintaining a fixed mapping relationship between the input and output optical fields of the MMF, i. e. with a stationary fiber shape. This is because the mode coupling phenomenon of the coherent light within the MMF will result in the highly environmentally sensitive output beam distribution [21-23]. The mapping relationship between the input and the output fields will be dynamically changing if the fiber shape changes, making focusing or imaging through the disturbed MMF almost impossible.

To eliminate the impact of fiber perturbations on reconstruction results, scientists propose to use the generalization ability of neural networks to enhance the robustness of imaging systems under dynamic changes in fibers [24-27]. In 2019, Fan et al. inserted a steel bar under the MMF to cause perturbation of the MMF and trained neural networks using speckles collected during the movement of the steel bar. They demonstrated that distorted images through a MMF subject to continuous shape variations could be recovered successfully [24]. However, the MMF used in this experiment was short, with a length of 35 cm, and the fiber perturbation caused by the movement of the steel rod was weak and less abrupt. Subsequently, Resisi et al. trained the network over hundreds of random nearly uncorrelated fiber bends, it succeeded in reconstructing high-fidelity images even when the fiber was strongly perturbed many weeks after the training perturbations [25]. Xu et al. proposed a U-architecture speckles imaging network (USINET), which had high robustness and learning invariance to the complex random variation of optical transmission characteristics inside the MMF [26]. However, neural networks require the training data must be independent and identically distributed with the test data [27]. Due to this limitation, a large number of input and output images of optical fibers in different shapes are required to achieve reconstruction, which will greatly increase the cost of training time. Without the support of a large number of samples, reconstruction after strong fiber perturbations becomes very difficult. Another possible method to eliminate the impact of fiber perturbations is to use newly developed glass-air Anderson localizing optical fibers (GALOFs) [28-30]. Hu et al. used GALOFs to achieve fiber imaging and observed an average 86.8% classification accuracy of cell images while the maximum bending offset distance is 45 cm [31]. Although the GALOFs seem promising in the fiber imaging systems, they are not Commercial-off-the-shelf available yet.

In addition to coherent light, incoherent light is also used as a light source for MMF imaging, with the advantages of simple installation and no coherent noise. In 2018, Shabairou et al. for the first time utilized light emitting diode (LED) as an incoherent illumination source in the MMF imaging system with a MMF length of 18 cm [32]. The light passing through the MMF can be reconstructed by an "autoencoder" network and a Sobel kernel filtering was implemented to filter the fiber output images in order to improve the image reconstruction effect. However, with increasing MMF length, the output beam distribution from the MMF will be more uniform because of the smaller spectral bandwidth [33], which will cause degradation of the filter performance, i. e. the image reconstruction capability. In 2022, Wang et al. compared the imaging effects of five light sources through MMFs and discovered that a broad incoherent light source was bad for image reconstruction due to that it was impossible to find an inverse transmission matrix that was suitable for all images and optical frequencies [34].

Although the previous reports show that a broadband incoherent light source may not be suitable for the MMF imaging system, in thin multimode fibers, the mode interference induced by incoherent light is weaker than that caused by coherent light, which can enhance the flexibility of imaging systems. LED is one of the cheapest and most readily available incoherent light sources. The key question is how to get a relatively good reconstruction effect when a broad bandwidth LED is used as the light source. In this paper, we propose a MMF imaging system based on white LED illumination and a reconstruction algorithm of a cascaded Unet combined with channel stitching technology. We reconstruct images of the Fashion-MINIST dataset with an average pearson correlation coefficient (PCC) of 0.83. We test the reconstruction effect of the Fashion-MINIST dataset through the MMF imaging system when the MMF undergoes bending, twisting, and fiber ring diameter changes. We also test the impact of changing the imaging distance on the reconstruction by changing the distance from the fiber input to the Digital Micro Mirror Device (DMD). At the same time, to simulate the situation of optical fibers entering complex environments, we completely change the fiber shape for reconstruction and find that the reconstruction effect is almost consistent with that of the initial shape. The experimental results demonstrate that our method has good robustness against fiber shape changes.

## 2. Principle

Under the illumination of coherent light, the mapping relationship between the input and output of the MMF can be described using a transmission matrix [14].

$$E_{OUT} = T_E E_{IN}, \quad (1)$$

where $E_{OUT}$ is the output light field, $T_E$ is the transmission matrix, and $E_{IN}$ is the input light field. Under the illumination of incoherent light, it can be changed to the transmission matrix of the light intensity field.

$$I_{OUT} = T_I I_{IN}, \quad (2)$$

where $I_{OUT}$ is the output light intensity distribution, $T_I$ is the transmission matrix, and $I_{IN}$ is the input light intensity distribution. At the same time, $I_{IN}$ is an image formed by an object passing through a fiber collimator on the end face of the fiber. Due to the use of light intensity to describe the transmission of incoherent light in multimode fibers, there is weak mode interference phenomenon within the fiber, so the perturbations of the fiber will cause weaker changes in the light intensity transmission matrix. The use of neural networks can ignore these changes and achieve the same reconstruction effect as before the perturbation.

White LED has a broad wavelength bandwidth, and the patterns formed by different bands of light reflected by the object passing through the MMF have different transmission matrices. Therefore, the output pattern can be described as:

$$I_{OUT} = \sum_{\lambda i=1}^{N} T_{\lambda i} I_{INi}, \quad (3)$$

where $I_{OUT}$ is the output light intensity distribution, $T_{\lambda i}$ is the transmission matrix of wavelength $\lambda i$, and $I_{INi}$ is the input light intensity distribution of wavelength $\lambda i$. In our experiment, $I_{INi}$ with different wavelengths is the same, which is the object loaded on the DMD. Therefore, different wavelengths of light transmitted through multimode fibers will form different patterns. To verify this, we illuminate the white light on one pixel-sized area on DMD (32 × 32 micromirrors) and use a color CCD to receive patterns output from the optical fiber. As shown in figure 1(a)-(d), the collected images can be divided into blue channel, green channel, and red channel. The structural similarity index measure (SSIM) between two pairs of each channel is tested, and the results are shown in figure 1(e). It is obvious that different channels of light output from the fiber have different patterns, which may be due to the different colors of the light reflected from the same point on the DMD exciting different modes inside the MMF. Therefore, we stitch patterns of different channels and input them into the neural network, enabling the network to obtain more features and improve the reconstruction effect.

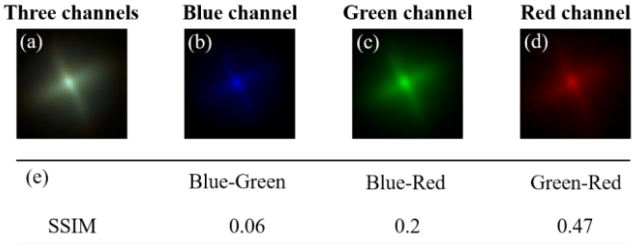

**Figure 1.** (a) The pattern received by a color CDD formed by the reflection of light from one pixel-sized area on the DMD passing through the MMF. (b)-(d) The patterns are obtained by dividing the three channels pattern into the blue channel, green channel, and red channel. (e) SSIM between each two channels.

## 3. Experiment setup

The experimental setup is shown in figure 2. The white light emitted by the LED (Daheng Optics, LED–20W) is directly illuminated on the intensity pattern loaded by the DMD. The DMD we used is UPOLabs HDSLM756D, which has a resolution of 1920 × 1080, a single micromirror size of 7.56 µm, and can achieve high-speed amplitude modulation at a refresh rate of about 9.2 kHz. Then the light reflected by the DMD is coupled into a 2 m long step-index MMF (62.5/125 µm, *NA* = 0.275) through a multimode fiber collimator (F1). The number of modes ranges from approximately 2500 to 9100. The object information is converted into the modes in the optical fiber for transmission. The pattern output from the optical fiber is collimated by a multimode collimator (F2) and then illuminated on the color CCD. The CCD we used is MER-132-43U3C-L of Daheng Optics, which has a resolution of 1292 × 964, a single pixel size length of 3.75 µm, and a frame rate of 43 fps. Finally, the patterns are saved in the computer by the frame capture from the CCD We process it into the red channel, green channel, and blue channel to obtain patterns. The computer is a Windows system with Intel i7-12700KF CPU and NVIDIA RTX 4090 GPU.

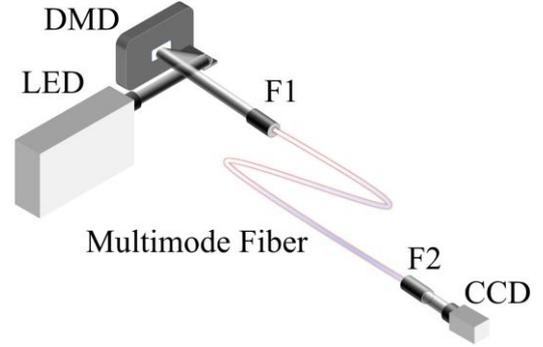

**Figure 2.** Experimental setup. The white light emitted by a LED illuminates the DMD. After the DMD reflection, it is coupled into the MMF through F1 (multimode fiber collimator). The light output from the MMF illuminates the CCD through F2 (multimode fiber collimator) and is recorded by the CCD.

## 4. Results

### 4.1 Image reconstruction

We utilize the 896 × 896 macro-pixel area of the DMD to simulate the object. White light illuminates the DMD, and its amplitude is modulated by the DMD to match the objects. After passing through F1, the object image is projected onto the surface of the DMD. Subsequently, the object image is converted into modes within the MMF for transmission. As a result, the output pattern carries information about the object, enabling the reconstruction of the object. We use the Fashion-MINIST dataset to test the system's reconstruction effect for grayscale images. The loaded image size is 28 × 28. Therefore, the resolution of the reconstructed image is 28 × 28. Since all images cover the entire fiber core area, the size of each pixel is 2.2 µm × 2.2 µm. To fit the input-output relationship of the MMF imaging system in this experiment, we adopt a cascaded Unet. Simply increasing the depth of the Unet to obtain features increases the number of parameters, while reconnecting a small Unet without skip connections can perform secondary filtering on features with a lower number of parameters. Therefore, we use an Unet to encode and decode the input image, and then pass through an Unet without skip connections for secondary encoding and decoding to improve the reconstruction effect. The structure of cascaded Unet is shown in figure 3(a). We collect 25680 images of 28 × 28 and their corresponding patterns of 128 × 128 as a training set. We use SSIM as the loss function and Adam as the optimizer. In order to utilize the band resources of white LED, we use color CCD to capture color patterns. At the same time, as shown in figure 3(b), we divide the collected patterns into three channels for grayscale processing and input them into the cascaded Unet for training through channel stitching technology. The trained model can directly reconstruct images, and the results are shown in figure 3(c). 3200 test sets are tested with an average PCC of 0.83 and an average SSIM of 0.61.

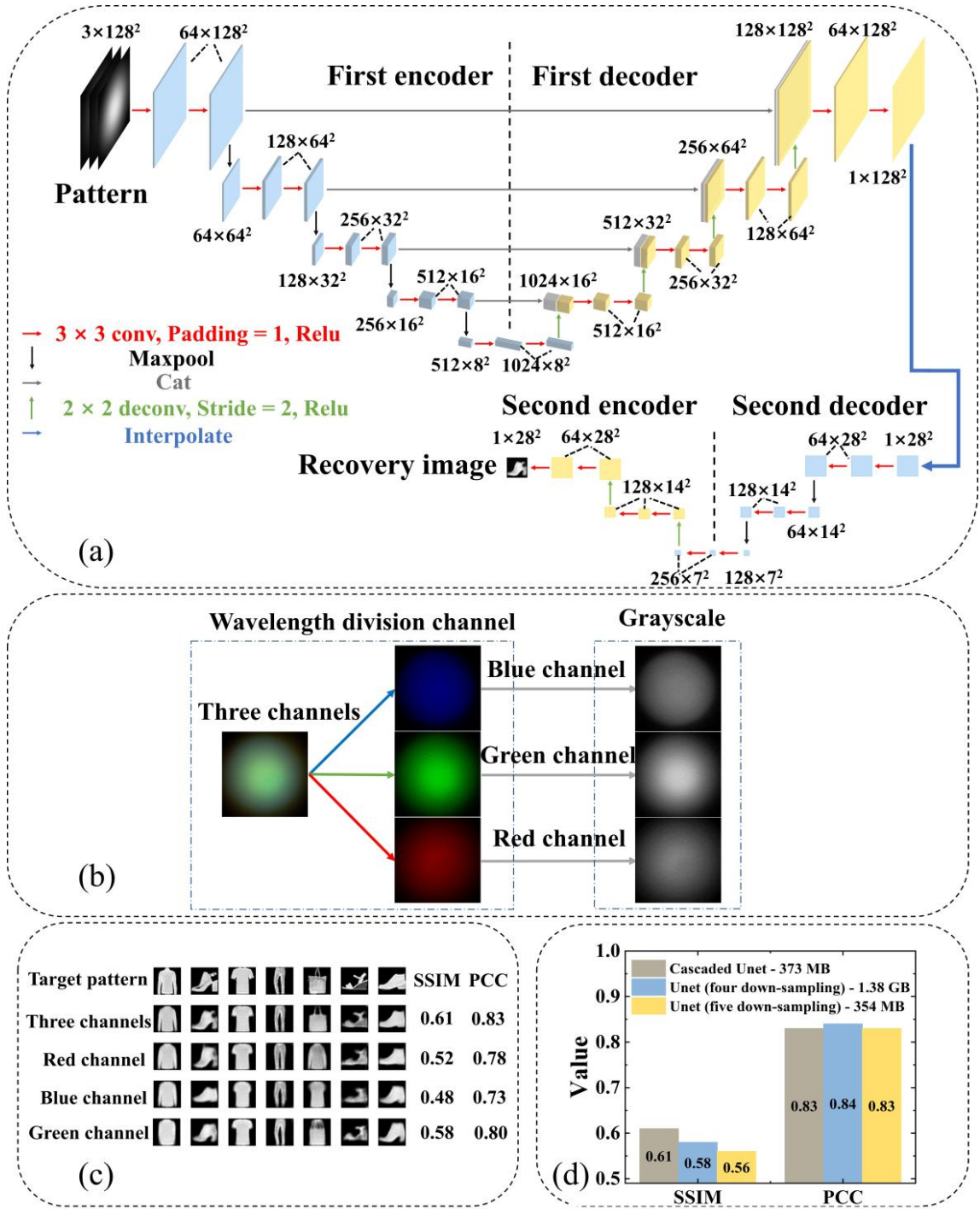

**Figure 3.** (a) Schematic diagram of the cascaded Unet structure. (b) Schematic diagram of three channels pattern processing. Divide the three channels pattern collected by the color CCD into blue channel, green channel, and red channel. Grayscale the images of each channel. Afterward, the images of the three channels are input into the cascaded Unet using channel stitching technology. (c) Reconstructed Fashion-MINIST dataset using different channels patterns and the average SSIM and PCC of 3200 test sets. (d) Comparison of the reconstruction effects of cascaded Unet, Unet with four down-sampling layers, and Unet with five down-sampling layers. All of them have the same training set, testing set, loss function, and optimizer. The Unet with four down-sampling layers has the same structure as the first Unet of cascaded Unet. The legend in the upper right corner also displays the sizes of the three networks.

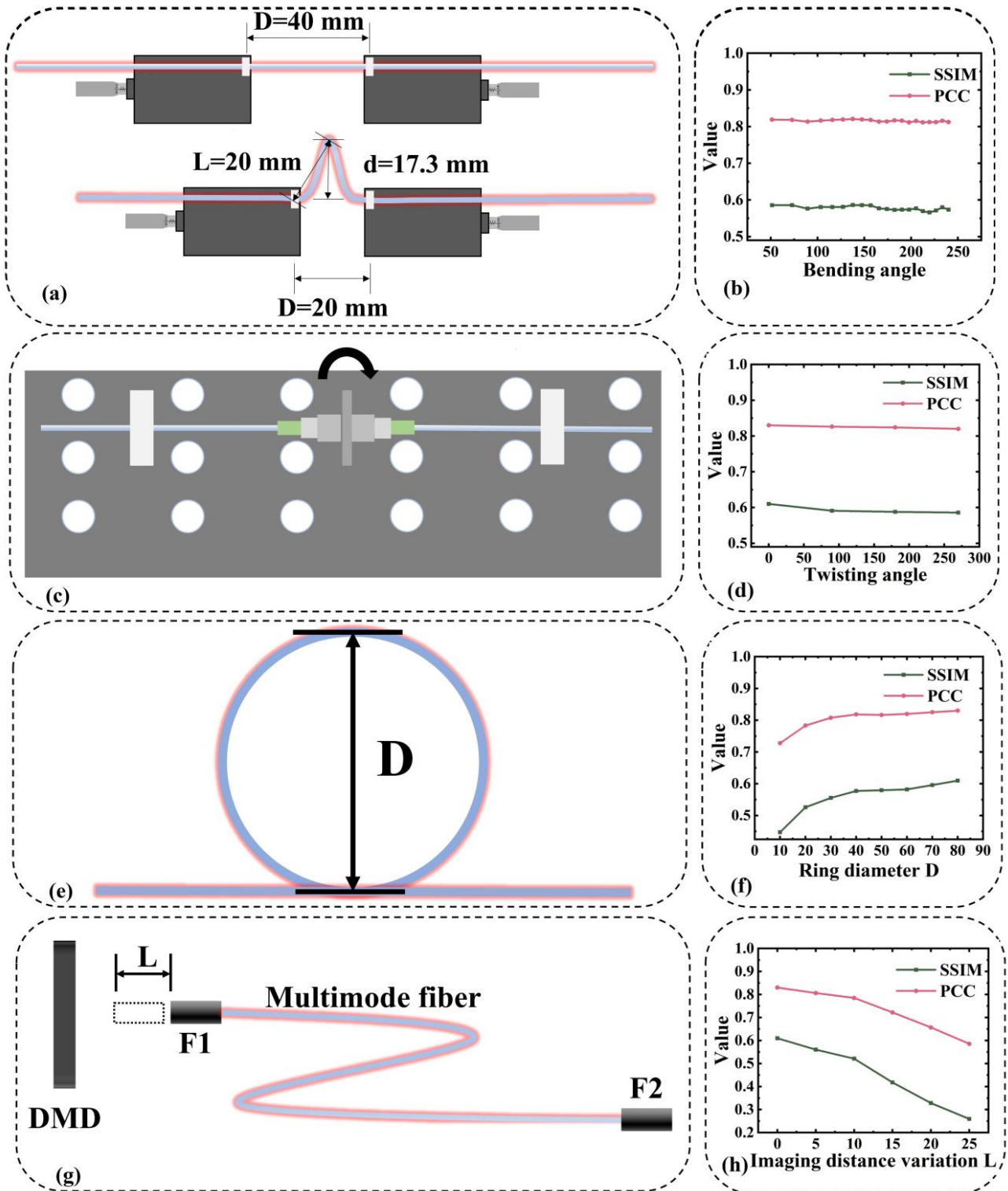

**Figure 4.** (a) Schematic diagram of fiber bending. Fix the straight section of the optical fiber on two displacement platforms, one of which is fixed, and use the other displacement platform to move the optical fiber to simulate the fiber bending at different angles. The straight section is 40 mm long, and the total displacement of the displacement platform is 20 mm. (b) The average SSIM and PCC of the reconstructed Fashion-MINIST dataset at different fiber bending angles. (c) Schematic diagram of fiber twisting. The optical fiber used in the experiment is composed of two 1-meter-long optical fibers connected by flange. Simulate fiber twisting by rotating the flange. The twisting angles are 90º, 180º, and 270º. (d) The average SSIM and PCC of the reconstructed Fashion-MINIST dataset at different twisting angles. (e) Schematic diagram of fiber ring diameter. Gradually reduce the diameter of the fiber ring from 80 mm to 10 mm. (f) The average SSIM and PCC of the reconstructed Fashion-MINIST dataset at different ring diameters. (g) Schematic diagram of imaging distance. Place the fiber input on the displacement platform, and change the imaging distance by controlling the displacement platform. Reduce the imaging distance by 25 mm from the initial distance. (h) The average SSIM and PCC of the reconstructed Fashion-MINIST dataset at different imaging distance variations.

To compare the reconstruction effect of single-band and multi-band images, we use the single red channel, blue channel, or green channel patterns to train the cascaded Unet. We test the reconstruction results of three neural networks under the same 3200 test sets, and their SSIM and PCC are shown in figure 3(c). The average PCC values are 0.78, 0.73, and 0.80 respectively. The average SSIM values are 0.52, 0.48, and 0.58 respectively. It can be seen that the use of channel stitching technology can improve the reconstruction ability of neural networks. This is because those different channels carry different information, and inputting three channels into a neural network can enhance the features of the input images, making it easier for the neural network to learn the mapping relationship between input light intensity distribution and output light intensity distribution. At the same time, the reconstruction effect of red and blue channels is worse compared to green channels. This is because that DMD is a diffractive device. When white light illuminates on the DMD, green light is located in the middle of the reflected light, while blue and red light are located on both sides of the green light. Due to aperture limitations in the fiber optic collimator, some blue and red light is blocked from entering the fiber, resulting in information loss. Therefore, the reconstruction effect is worse compared to the green light.

In order to verify the superiority of cascaded Unet in this experimental task, we compare it with traditional Unet, which has the same structure as the first Unet of cascaded Unet. The results are shown in figure 3(d). We train the traditional Unet with the same training set and test it using the same 3200 images. The average PCC and SSIM of the reconstructed images are 0.83 and 0.56. It can be seen that the cascaded Unet has better reconstruction ability compared to the Unet with four down-sampling layers, which proves the feasibility of repeatedly extracting features to improve reconstruction results. We also test the reconstruction effect of an Unet with five down-sampling layers. The average PCC and SSIM of the reconstructed images are 0.84 and 0.58. It is found that increasing the depth of the neural network can improve the reconstruction effect, but it will add a large number of parameters. Compared to cascaded Unet, although Unet with five down-sampling layers has a higher PCC of reconstructed images, its SSIM of reconstructed images is lower and its parameters are 3.8 times larger than that of cascaded Unet. Hence, we connect a small Unet after an Unet in this paper to repeatedly extract features and improve the reconstruction effect with a small number of additional parameters.

## 4.2 Perturbation testing

In order to demonstrate the system's ability to resist fiber perturbations, we test the effect of different fiber bending angles on reconstruction results. As shown in figure 4(a), we add two displacement platforms to the straight section of the optical fiber. One of the platforms is fixed, and the other displacement platform is used to squeeze the fiber, simulating a bending to the fiber. The straight section of the optical fiber is 40 mm long. Using the displacement platform to move the fiber of 20 mm, calculate the fiber lateral displacement of 17.3 mm and the bending angle of 240º. We measure the SSIM and PCC of the reconstructed images within 20 mm of the displacement platform moving in steps of 1 mm. The results are shown in figure 4(b). It can be seen that as the displacement increases, the SSIM and PCC of the reconstructed images remain above 0.56 and 0.80. The experimental result indicates that our method still has good reconstruction results even with increasing fiber bending angle.

At the same time, we measure the impact of the fiber twisting on the image reconstruction. Our MMF is composed of two 1-meter-long optical fibers connected through flanges. As shown in figure 4(c), we fix the optical fibers at both ends of the flange and simulate the fiber twisting by 90º, 180º, and 270º by rotating the flange. We test the average SSIM and PCC for reconstructed images under three different twisting angles, and the results are shown in figure 4(d). It can be seen that whether it is SSIM or PCC, the reconstruction effect after twisting is close to that without twisting.

When light passes through a curved section inside a MMF, there will be modes leak. To test the impact of fiber ring diameter on the image reconstruction effect, as shown in figure 4(e), we initially fix an 80 mm diameter fiber ring and reduce the diameter in steps of 10 mm. We measure the average SSIM and PCC of the reconstructed images under different ring diameters and show them in figure 4(f). It can be seen that as the diameter of the ring decreases, the reconstruction effect becomes worse. This is because reducing the diameter of the circular ring will cause modes leak, which will cause changes in the fiber output and make it difficult for the neural network to reconstruct. However, as long as the diameter of the fiber ring is larger than 20 mm, our system still has good reconstruction performance.

To test the effect of imaging distance on image reconstruction, as shown in figure 4(g), we use a displacement platform to pull in the distance between the fiber input and the DMD. The average SSIM and PCC of reconstructed images are measured at displacements of 5 mm, 10 mm, 15 mm, 20 mm, and 25 mm, and the results are shown in figure 4(h). It can be seen that as the initial position of the fiber input principle moves, the image reconstruction effect gradually deteriorates, but it can still reconstruct the image well even at a displacement of 10 mm. Therefore, we can achieve reconstruction with a coupling distance error of 10 mm, and in the actual imaging process, there is no need to strictly maintain the distance between the object and the fiber input, reducing the difficulty of imaging.

In practical applications, the shape of optical fibers can undergo drastic changes. To simulate this process, we thoroughly change the initial shape of the optical fiber, as shown in figure 5(a). We test the reconstruction effect after perturbation as shown in figure 5(b). The results show that even with the overall change of the MMF, the reconstruction effect can still be completely consistent with the initial shape. Therefore, using the white LED as the light source of the MMF imaging system has strong robustness.

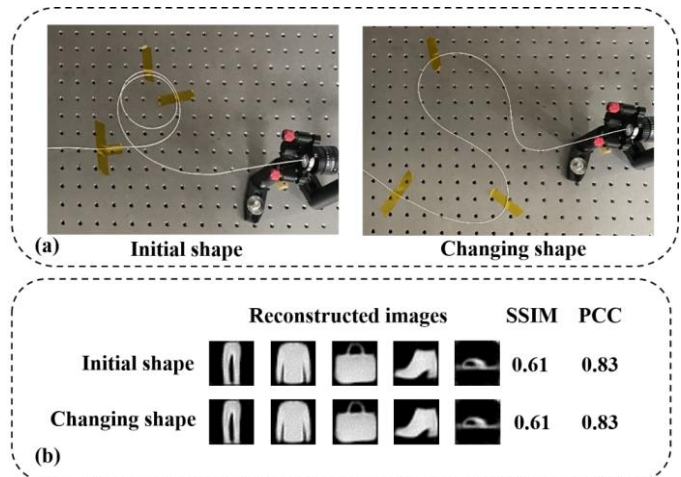

**Figure 5.** (a) Schematic diagram of the initial shape of the optical fiber and the changing shape of the optical fiber. (b) Reconstruction results of the Fashion-MINIST dataset at the initial shape and changing shape.

## 5. Conclusion

In this article, we successfully utilize white LED and cascaded Unet to achieve MMF imaging and achieve almost the same reconstruction effect as that of the initial shape when completely changing the fiber shape. The average PCC reconstructed from the Fashion-MINIST dataset in the initial shape is 0.83. At the same time, we successfully improve the SSIM and PCC of the reconstructed images using image channel stitching technology. Compared with using only single-channel patterns for image reconstruction, patterns in RGB channels processed by channel stitching technology have richer features, which is conducive to neural networks finding the mapping relationship between the input and output optical fields. The cascaded Unet we utilized has a better reconstruction effect compared to the Unet with four down-sampling layers and a lower number of parameters compared to the Unet with five down-sampling layers. We prove the feasibility of repeatedly extracting features instead of deepening the structure to extract features. Compared to coherent light sources, our use of LED as the light source has stronger robustness against fiber shape changes. At the same time, our method does not require strict control of the imaging distance between the optical fiber and the object, which can move freely within 10 mm. Our method provides a new idea for resisting fiber perturbations and contributes to the development of MMF imaging for practical applications.

## Acknowledgments

This work was supported by the National Natural Science Foundation of China (Grant No. 61975137). The authors declare no conflicts of interest.